# PENUMBRAL WAVES DRIVING SOLAR FAN-SHAPED CHROMOSPHERIC JETS

A. Reid[1], Henriques, V. M. J.[2,3,1], M. Mathioudakis[1], T. Samanta[4]
1. Astrophysics Research Centre, School of Mathematics and Physics, Queen's University Belfast, BT7 1NN, Northern Ireland, UK; e-mail: aaron.reid@qub.ac.uk
2. Institute of Theoretical Astrophysics, University of Oslo, PO Box 1029 Blindern, 0315 Oslo, Norway
3. Rosseland Centre for Solar Physics, University of Oslo, P.O. Box 1029 Blindern, NO-0315 Oslo, Norway
4. School of Earth and Space Sciences, Peking University, Beijing 100871, China

## ABSTRACT

We use H$\alpha$ imaging spectroscopy taken via the Swedish 1-m Solar Telescope (SST) to investigate the occurrence of fan-shaped jets at the solar limb. We show evidence for near-simultaneous photospheric reconnection at a sunspot edge leading to the jets appearance, with upward velocities of 30 km s$^{-1}$, and extensions up to 8 Mm. The brightening at the base of the jets appears recurrent, with a periodicity matching that of the nearby sunspot penumbra, implying running penumbral waves could be the driver of the jets. The jets' constant extension velocity implies that a driver counteracting solar gravity exists, possibly as a result of the recurrent reconnection erupting material into the chromosphere. These jets also show signatures in higher temperature lines captured from the Solar Dynamics Observatory (SDO), indicating a very hot jet front, leaving behind optically thick cool plasma in its wake.

*Keywords:* Sun: Activity — Sun: Chromosphere — Sun: Photosphere — Sun: Oscillations

## 1. INTRODUCTION

Spicules and fibrils are well documented examples of chromospheric plasma motions appearing as thin, jet-like strands. Type I spicules and dynamic fibrils are thought to be driven by magneto-acoustic shocks formed by p-modes, while type-II spicules are thought to be driven by reconnection events (Hansteen et al. 2006; Tsiropoula et al. 2012) or via the release of magnetic tension (Martínez-Sykora et al. 2017).

Fan-shaped jets (also known as peacock jets (Robustini et al. 2016) and light walls (Yang et al. 2015)) have been subject to much less study due to the rarity of observing such events. These jets were first reported by Roy (1973), who described them as solar surges appearing in the centres of active regions in the H$\alpha$ line. They reported strong initial acceleration with peak velocities of 150 km s$^{-1}$ and extensions up to 50 Mm. Deceleration due to gravity eventually caused these surges to reverse direction and fall back towards the solar surface. Until the advent of high resolution ground-based observatories, very little was added to this description. Asai et al. (2001) combined ground-based and space-borne instruments to note that fan-shaped jets were recurrent, and appear dark in the blue wings of the H$\alpha$ line, and in the Fe IX 171 Å line, normally sensitive to transition region/coronal temperatures. They also found less extreme velocities of ~40 km s$^{-1}$, lengths of 15 - 20 Mm, and mean lifetimes of 10 minutes.

On-disk fan-shaped jets most often appear at light bridges (Roy 1973; Asai et al. 2001; Robustini et al. 2016), where the dominant magnetic field orientation in the photosphere changes from vertical to horizontal (Shimizu et al. 2009). This has led to the theory that fan-shaped jets are formed via shearing reconnection between the vertical and horizontal photospheric magnetic field lines at the light bridge - umbra interface (Robustini et al. 2017). Recently, Hou et al. (2016) reported a solar flare occurring along the same field lines as fan-shaped jets, causing them to incline and shorten.

Three-dimensional simulations have reinforced this theory, showing fan-shaped jets being formed during shearing reconnection between horizontal magnetic fields and a vertical current sheet, with the base of the jets corresponding to the location of magnetic reconnection (Jiang et al. 2011). While the initial acceleration of the jets is due to the magnetic tension, the kinematics following this are thought to be due to the gas pressure gradient created via the heating at the base of the jets (Li et al. 2016).

Yang et al. (2016) noted the bright fronts of the jets were coupled to recurrent brightenings at the jets' base, and that these brightenings resulted in lengthening of the jets. This fits with the theory of recurrent photospheric reconnection causing this phenomenon. Zhang et al. (2017) suggest that these bright fronts are caused by propagating shocks due to p-modes leaking from the photospheric base.

On-disk fan-shaped jets appear dark in the wings of the H$\alpha$ line, while appearing bright in the line cores. They also appear as brightenings in the Ca II H (Shimizu et al. 2009) and Ca II 8542 Å lines (Robustini et al. 2017), while the fronts appear bright in the *Solar Dynamics Observatory (SDO) Atmospheric Imaging Assembly* (AIA, Lemen *et al.* 2012), 304 Å and 171 Å channels (Robustini et al. 2016), and the *Interface Region Imaging Spectrograph* (IRIS, De Pontieu et al. 2014) 1330 Å channel (Yang et al. 2016; Tian et al. 2018).

In this paper we report on observations of recurrent photospheric brightenings at a sunspot edge, resulting in the production of fan-shaped jets. We describe the observational setup in Section 2, while we discuss the appearance and dynamics of these jets in Section 3. Section 4 draws conclusions from these observations.

## 2. OBSERVATIONS

The observational data were obtained with the CRisp Imaging SpectroPolarimeter (CRISP) at the Swedish 1-m Solar Telescope (SST, Scharmer *et al.* 2003; Scharmer *et al.* 2008) on La Palma. The target was an active region (NOAA 12126) at the limb (coordinates: X= 897″, Y= -184″, $\mu$ = 0.34). The observations took place on 2014 August 02 between 09:07 and 09:54 UT. The observations comprised of H$\alpha$ imaging spectroscopy. Each scan sampled 11 line positions, taken at

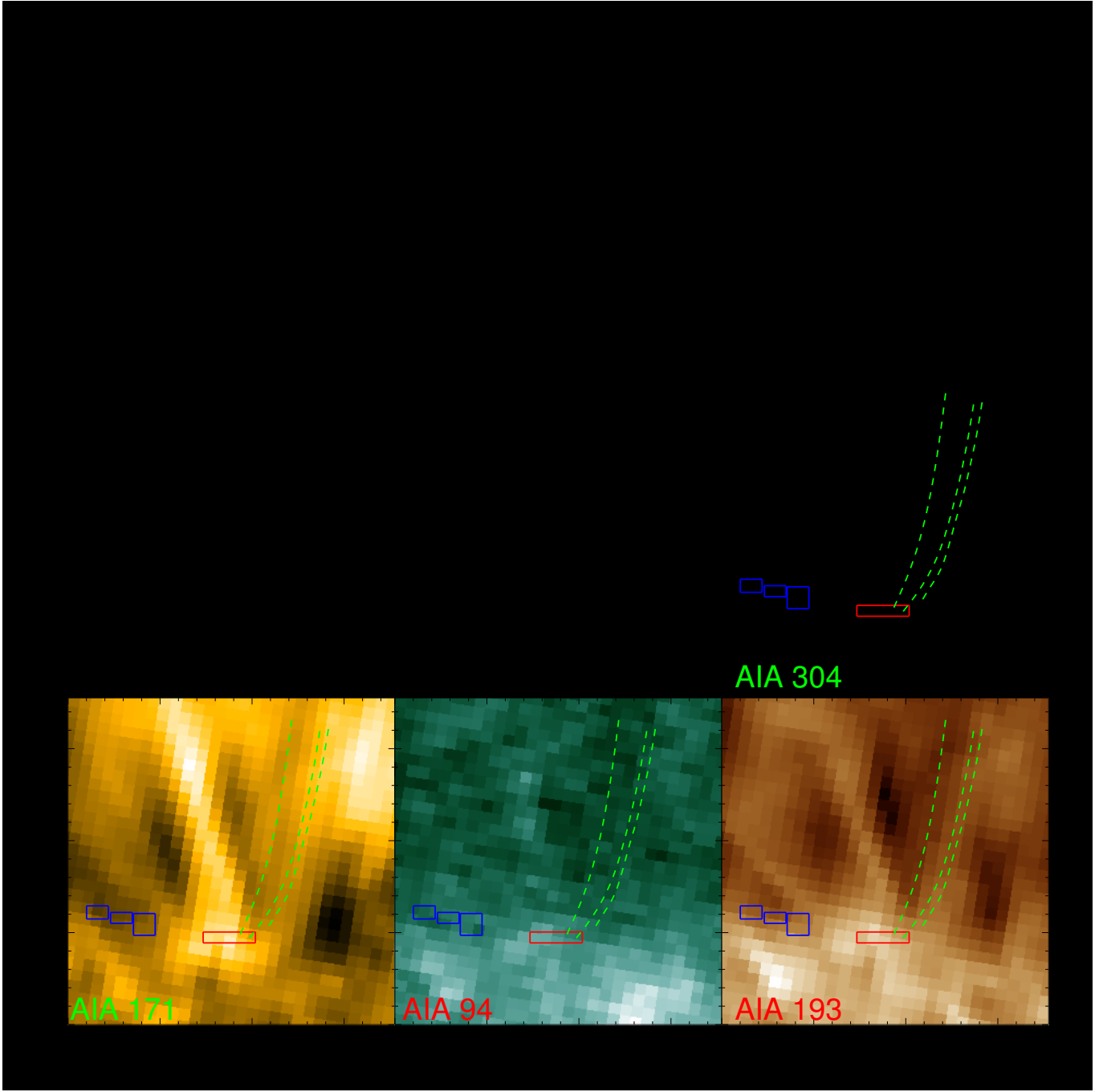

**Figure 1.** Region-of-interest (ROI) showing fan-shaped jets with co-aligned SDO AIA data, with dimensions 17.7″x 17.7″. Top: Hα - 0.814 Å, +0.000 Å, +0.814 Å. Middle: AIA 1600 Å, 1700 Å, 304 Å. Bottom: AIA 171 Å, 94 Å, 193 Å. The green lines shows curvilinear slices of example jets. The red-box shows the brightened region, while the blue boxes indicate the regions used to measure the penumbra. A movie is also available. (see Movie1)

± 1.450 Å, ± 1.087 Å, ± 0.815 Å. ± 0.543 Å, ± 0.317 Å, and line center. The Hα data had a post-reduction mean cadence of 3 seconds, and a pixel scale of 0.059″/pix.

The SST data were processed with the Multi-Object Multi-Frame Blind Deconvolution (MOMFBD) algorithm (Löfdahl 2002; van Noort et al. 2005). This includes tessellation of the images into 88x88 pixels$^2$ sub-images for individual restoration to preserve the assumption of invariance of the image formation models. An extended MOMFBD scheme that includes the reconstruction of auxiliary wide-band images, co-temporal with the narrow-band wavelengths were used to-gether with de-stretching (Shine et al. 1994) to reduce the impact of residual seeing on the profiles (Henriques 2012). The field-of-view (FOV) of the reconstructed dataset is 59″x 58″. Prefilter field-of-view and wavelength dependent corrections were applied to the restored images. More information relating to the SST reduction pipeline used in this manuscript is available in de la Cruz Rodríguez et al. (2015).

Co-aligned SDO AIA and Helioseismic and Magnetic Imager (HMI, Schou et al. 2012) data were created using Rob Rutten's alignment software, available on his website [1]. The region-of-interest (ROI) is shown in Figure 1, and consists of a sunspot, with a nearby pore also visible. The fan-shaped jets appear between the umbra of the main sunspot and the pore. Throughout the time-series, coronal rain can be seen in the background, along with numerous Ellerman Bombs erupting in the surrounding photosphere. The foot-point of an overlying loop structure is visible in the high temperature AIA 171 Å and AIA 193 Å images.

---

[1] http://www.staff.science.uu.nl/ rutte101/rridl/

### 3. RESULTS AND ANALYSIS

The jets can be clearly seen as brightenings in the H$\alpha$ core relative to the background, while appearing dark in the H$\alpha$ red wing, and less visible in the H$\alpha$ blue wing. This asymmetry is most likely due to the orientation of the jets relative to the line-of-sight. The jets are also visible in the AIA 304 Å and 171 Å channels, with the jet tips appearing bright, while they are not visible in the low temperature 1600 Å and 1700 Å filters. The body of the jets appear as dark structures in the AIA 171 Å, 193 Å, and 131 Å channels. This darkening is also visible in 335 Å, 211 Å, and 94 Å with very poor contrast.

In the photospheric H$\alpha$ wings, AIA 1600 Å & 1700 Å, a brightening is apparent at the edge of the sunspot penumbra and at the base of the jets. In a similar fashion to the jets, this photospheric brightening is recurrent, possibly due to magnetic flux being continuously supplied by the decaying sunspot (Kubo et al. 2008). The recurrence of the jets can be seen in the H$\alpha$ core image of Figure 1, showing a bright front, which moves up the existing jet structure.

We constructed light-curves for the photospheric brightening at the base of the jets by integrating over the red box of Figure 1. The resultant light-curves are shown in Figure 2. The recurrent nature of the photospheric signals in H$\alpha$, 1600 Å, and 1700 Å prompted us to compute wavelet power transforms, which are shown in Figure 3. The cone of influence (COI) is calculated based on a significance level of 95%. Strong power is apparent in the brightening with a periodicity of 275 seconds. The wavelet analysis of Figure 3 shows the results using the AIA 1700 Å channel which had the strongest periodicity, however the same peak power was found in the H$\alpha$ and AIA 1600 Å channels. This peak period was not apparent in the other AIA passbands. Periodic sine waves have been added to the light-curves in Figure 2, showcasing the agreement between the periodicity and the recurrent photospheric brightening.

Wavelet analysis was also done for the local sunspot penumbra and the nearby quiet-Sun reference profiles, and are shown in Figure 3. The regions taken for the penumbral light-curves are shown in Figure 1. These are as close as possible to the brightening without interference from overlying Ellerman Bombs or the brightening itself. The quiet-Sun region was taken outside of the sub-FOV of Figure 1. The penumbra wavelet power transform also shows the same peak periodicity of 275 seconds, while this is not apparent in the reference profile.

The periodicity of the penumbra is similar to that calculated previously for running penumbral waves (Löhner-Böttcher & Bello González 2015), and could be the same phenomenon. This suggests that the recurrent brightening at the sunspot edge could be driven by the same running penumbral waves, pushing outwards to the sunspot edge.

While the brightening appeared repetitive, between peak period times material was observed to erupt upward into the chromosphere. This is seen best in Figure 4, where the upflows captured in the blue wing of H$\alpha$ show how the photospheric brightening pushes material into the chromospheric jets.

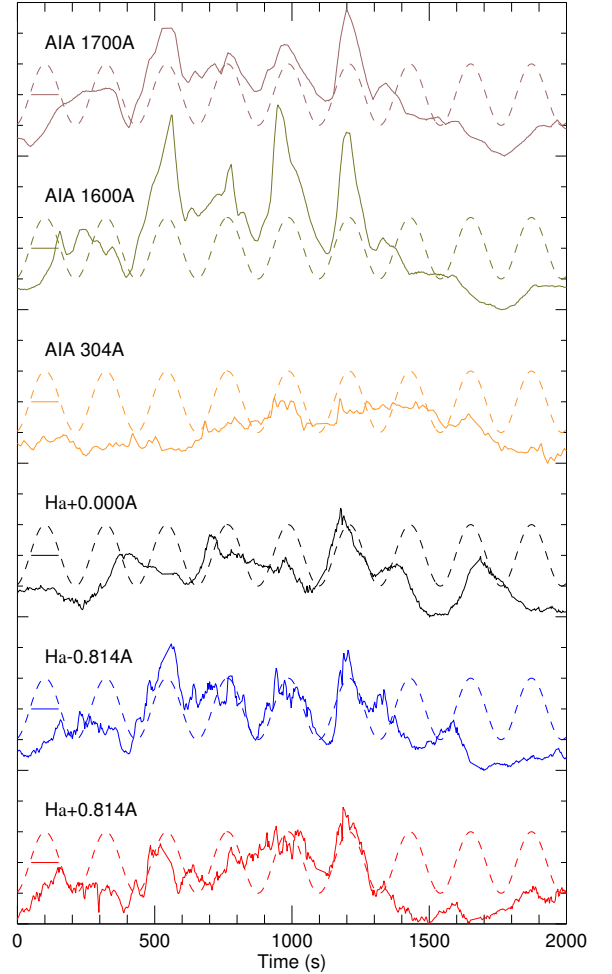

**Figure 2.** Light-curves of the recurrent photospheric brightening (solid), averaged from the red-box in Figure 1. Periodic sine waves are also created based on the peak period from the global wavelet transforms of Figure 3 (dashed).

Three of the fine jet structures are highlighted by the dashed lines in Figure 1. Curvilinear time-slices were then created for these jets. The time-slice for the right-hand jet is shown in Figure 5, which best highlights the visibility of these jets across the various wavelengths. The peak in the photospheric light-curves occurs simultaneously to the appearance of the traced jet front at T = 1195 seconds in H$\alpha$ line core and SDO AIA channels. At T = 1420 seconds, the traced jet reaches its maximum protrusion of 7.6 Mm. The repeated photospheric intensity enhancement always peaks near-simultaneously to the emergence of a new bright jet front.

While the emanation of the jet fronts appear identical in the H$\alpha$ line core and blue wing, in the H$\alpha$ red wing, a contraction can be seen. The H$\alpha$ red wing appears to show the falling of jets back to the solar surface rather than the presence of new bright jets, indicating a line-of-sight effect and that the jets are slightly tilted towards the observer. Doppler velocities have been estimated via fitting of a double gaussian to the line profiles. However, our spectral sampling limits the Doppler velocity resolution to ±14 km s$^{-1}$. The falling jets appeared to have a Doppler shift of ∼30 km s$^{-1}$, while the emanating jets do not appear to show any Doppler motion. This implies the emanating jets are propagating along the plane-of-sky, while the falling jets are slightly angled towards the observer (up to 20 degrees from the plane-of-sky). This could be due to the emanating jets traveling along the orientation of the field lines (Jiang et al. 2011), while the falling jets will be falling due to gravity. Using the Hα time-slices, the jet front appears to propagate upwards at a consistent speed of 28 - 30 km s$^{-1}$, and does not appear to decelerate until almost at peak length. The apparent falling of the same jet front in the red wing of Hα however appears to show a constant deceleration due to solar gravity of ∼0.3 km s$^{-2}$, reaching down-flow velocities > 100 km s$^{-1}$. The blue dashed line of Figure 5 shows the polynomial fitting of the falling material, while the green-dashed line shows the linear fitting of the rising jet front. The pink dotted line shows the manual trace of the propagating jet front. Both the quadratic and linear functions appear to fit the jet front, and so $\chi^2$ values were calculated. The $\chi^2$ values of the linear and polynomial fitting against the rising jet front are 15.36 and 947.15 respectively, validating that the rising jet front appears to emanate with a constant velocity.

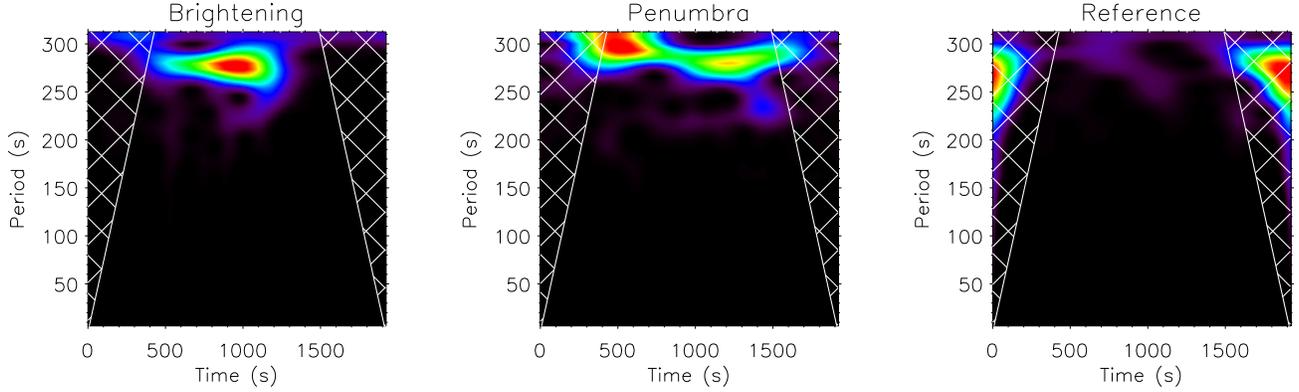

**Figure 3.** Wavelet power transforms computed for the SDO 1700 Å channel. The crossed regions show regions in the diagrams outside of the cone of influence. Left: The brightening (red box in Figure 1). Middle: The local sunspot penumbra. Right: Nearby quiet-Sun reference.

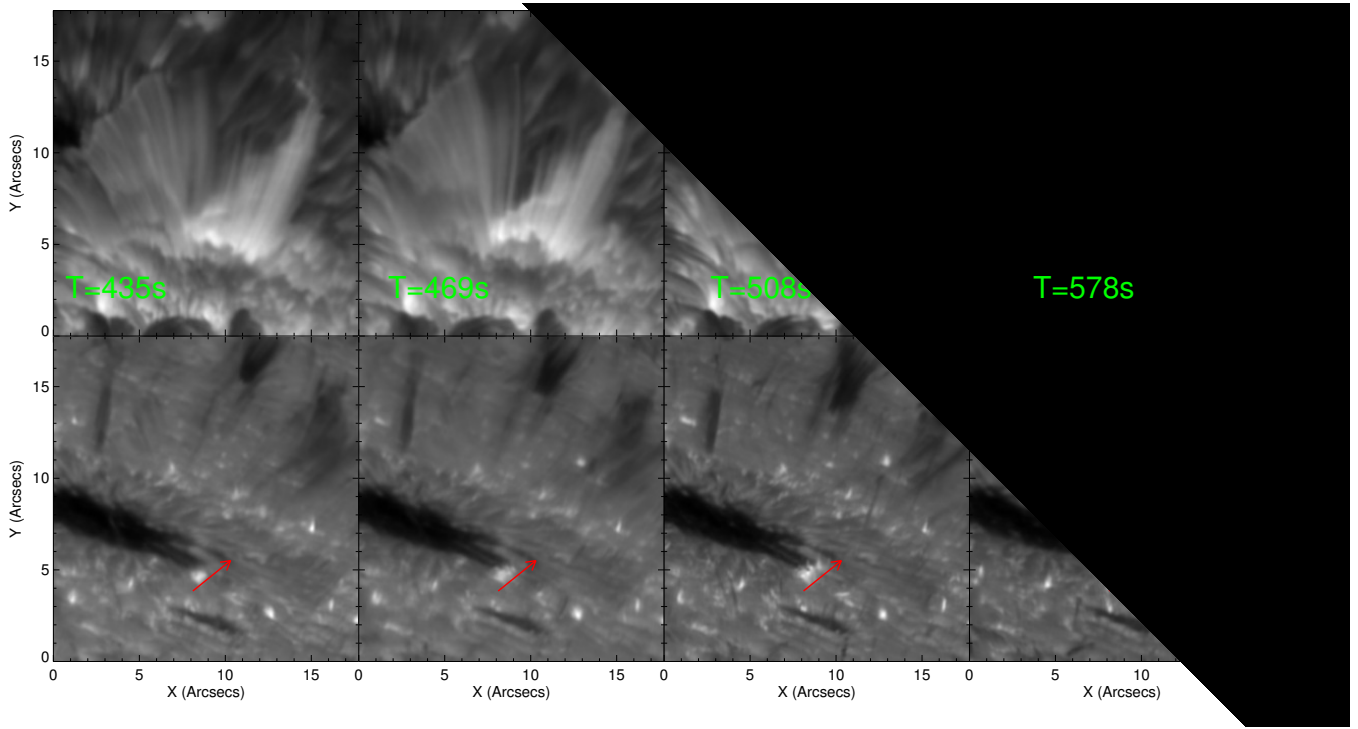

**Figure 4.** Top: Hα line core images showing the progression of the jet fronts. Bottom: Co-temporal Hα -0.814 Å images showing the eruption of material from the photospheric brightening into the jets. The scale is identical to Figure 1. The red arrow shows the direction of bright material erupting into the chromosphere (left of the arrow).

## 4. DISCUSSION AND CONCLUSIONS

We have investigated recurrent fan-shaped jets at the solar limb. The formation of the fan-shaped jets occur near simultaneously with the occurrence of a corresponding photospheric brightening. The peak frequency of the recurrent photospheric brightening matches the peak frequency of the penumbra of the nearby large sunspot, suggesting that run-ning penumbral waves could be triggering this form of reconnection.

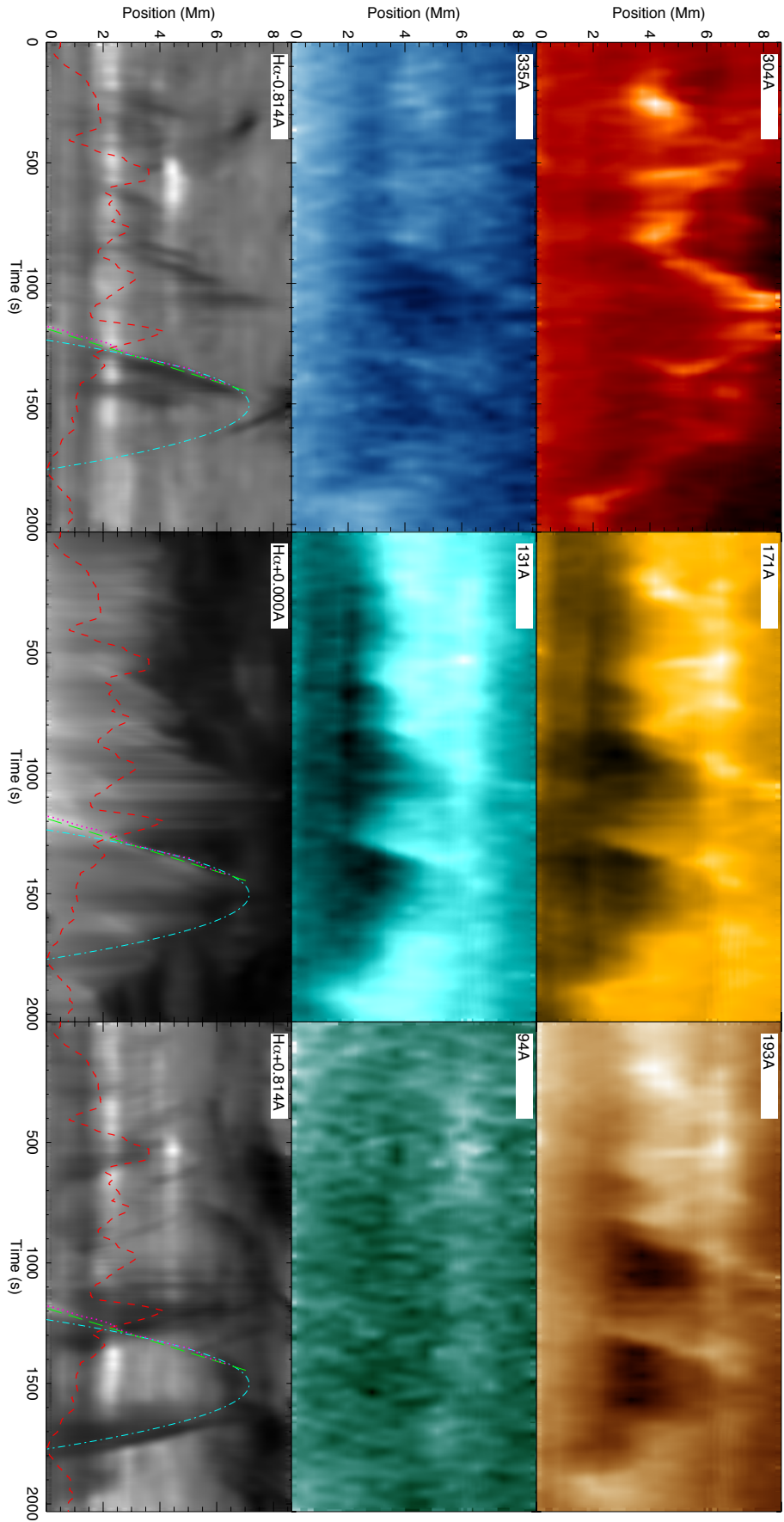

**Figure 5.** Time-slices of the green dashed line in Figure 1. The red dashed line indicates the 1700 Å light-curve from Figure 2. The blue dashed line indicates the polynomial fitting of the falling material due to gravity, while the green dashed line shows the linear fitting of the rising velocity. The dotted purple line shows the manual fitting of the jet front.

The peak extensions of the jets (∼7-8 Mm) and their apparent peak velocities are shorter than those reported for fan-shaped jets over light-bridges. However, the acceleration of the jets back to the solar surface does match with previous deceleration and velocity measurements. One possible explanation for the jets maintaining a steady extension velocity is that they are being constantly driven by the eruptive processes at the base which counteracts the effects of solar gravity, rather than emanating with a fast (∼100 km s$^{-1}$) burst which decelerates due to gravity.

The jets appear with bright fronts in the SDO AIA 171 Å and 304 Å channels, while in the passbands sensitive to much higher temperatures (including 171 Å), a darkness is left in the wake of the protruding jet front. This darkening is most likely optically thick plasma caused by strong absorption due to an increased density of hydrogen and singly ionised helium (Anzer & Heinzel 2005). This will absorb any emission from the higher energy Fe lines from behind the event, along the same line-of-sight.

The Swedish 1-m Solar Telescope is operated on the island of La Palma by the Institute for Solar Physics of Stockholm University in the Spanish Observatorio del Roque de los Muchachos of the Instituto de Astrofísica de Canarias. The research leading to these results has received funding from the European Community's Seventh Framework Programme (FP7/2007-2013) under grant agreement no. 606862 (F-CHROMA). V. Henriques was supported by the Research Council of Norway through its Centres of Excellence scheme, project number 262622, and by the Norwegian Research Council (project 250810 / F20). This work is supported by NSFC under grant 41574166, the Recruitment Program of Global Experts of China, and the Max-Planck Partner Group program. AR would like to Krishna Prasad for valuable comments, and the anonymous referee for providing valuable insight which improved this manuscript.

## REFERENCES

Anzer, U., & Heinzel, P. 2005, ApJ, 622, 714
Asai, A., Ishii, T. T., & Kurokawa, H. 2001, ApJ, 555, L65
de la Cruz Rodríguez, J., Löfdahl, M., Sütterlin, P., Hillberg, T., Rouppe van der Voort, L. 2015, A&A 573, A40
De Pontieu, B., Title, A. M., Lemen, J. R., et al. 2014, Sol. Phys., 289, 2733
Hansteen, V. H., De Pontieu, B., Rouppe van der Voort, L., van Noort, M., & Carlsson, M. 2006, ApJ, 647, L73
Henriques, V. M. J. 2012, A&A, 548, A114
Hou, Y., Zhang, J., Li, T., et al. 2016, ApJ, 829, L29
Jess, D. B., Reznikova, V. E., Van Doorsselaere, T., Keys, P. H., & Mackay, D. H. 2013, ApJ, 779, 168
Jiang, R.-L., Shibata, K., Isobe, H., & Fang, C. 2011, Research in Astronomy and Astrophysics, 11, 701
Kubo, M., Lites, B. W., Ichimoto, K., et al. 2008, ApJ, 681, 1677-1687
Lemen, J. R., Title, A. M., Akin, D. J., et al. 2012, SoPh, 275, 17
Li, Z., Fang, C., Guo, Y., et al. 2016, ApJ, 826, 217
Löfdahl, M.G. 2002, SPIE, 4792, 146L
Löhner-Böttcher, J., & Bello González, N. 2015, A&A, 580, A53
Martínez-Sykora, J., De Pontieu, B., Hansteen, V. H., et al. 2017, Science, 356, 1269
Robustini, C., Leenaarts, J., de la Cruz Rodriguez, J., & Rouppe van der Voort, L. 2016, A&A, 590, A57
Robustini, C., Leenaarts, J., & de la Cruz Rodríguez, J. 2017, arXiv:1709.03864
Roy, J.-R. 1973, Sol. Phys., 32, 139
Scharmer, G. B., Bjelksjo, K., Korhonen, T. K., Lindberg, B., & Petterson, B. 2003a, Society of Photo-Optical Instrumentation Engineers (SPIE) Conference Series, Vol. 4853, ed. S. L. Keil & S. V. Avakyan, 341-350
Scharmer, G. B., Narayan, G., Hillberg, T., de la Cruz Rodríguez, J., Löfdahl, M. G., Kiselman, D., Sutterlin, P., van Noort, M. & Lagg, A. 2008, ApJ 689, L69
Schou, J., Scherrer, P. H., Bush, R. I., et al. 2012, SoPh, 275, 229
Shimizu, T., Katsukawa, Y., Kubo, M., et al. 2009, ApJ, 696, L66
Shine, R. A., Title, A. M., Tarbell, T. D., Smith, K., Frank, Z. A., Scharmer, G. 1994, ApJ, 430, 413
Tian, H., Yurchyshyn, V., Peter, H., et al. 2018, ApJ, 854, 92
Tsiropoula, G., Tziotziou, K., Kontogiannis, I., et al. 2012, Space Sci. Rev., 169, 181
van Noort, M., Rouppe van der Voort, L. & Löfdahl, M. G. 2005, SoPh, 228, 191
Yang, S., Zhang, J., Jiang, F., & Xiang, Y. 2015, ApJ, 804, L27
Yang, S., Zhang, J., & Erdélyi, R. 2016, ApJ, 833, L18
Zhang, J., Tian, H., He, J., & Wang, L. 2017, ApJ, 838, 2